\documentclass[prl,twocolumn,amsmath,amssymb]{revtex4}

\usepackage{color}
\usepackage{bm}
\usepackage{dcolumn}
\usepackage[dvips]{graphicx}
\usepackage{multirow}

\usepackage{ulem}

\newcommand{\cin}[1]{{\color{black}#1}}
\newcommand{\cinn}[1]{{\color{black}#1}}
\newcommand{\cinb}[1]{{\color{black}#1}}

\begin{document}

\preprint{preprint}

\title{Diophantine equation for the Rice--Mele model:\\
Topological aspect of filling numbers and associated spatial pump
}

\author{Koichi Asaga}
\author{Takahiro Fukui}
\affiliation{Department of Physics, Ibaraki University, Mito 310-8512, Japan}

\date{\today}

\begin{abstract}
We introduce a long-period generic spatial modulation 
into a typical model of the Thouless pump, namely, the Rice--Mele (RM) model,
to examine the lattice analog of the fermion charge in quantum field theory.
We derive a Diophantine equation 
relating the fermion charge and the pumped charge, which leads to 
the one-dimensional (1D) analog of the Streda formula in the quantum Hall effect (QHE).
This formula 
implies that an adiabatic change of the periodicity of the spatial modulation yields a spatial charge pump 
such that  the rightmost charge is pumped to the right by the Chern number compared with the leftmost charge.
This causes a change 
in the length of the fermion chain by an integer, 
thus providing the opportunity for direct measurement of 
the  Streda formula in 1D systems.
\end{abstract}


\maketitle

The discovery of solitons in polyacetylene by Su, Schrieffer, and Heeger had a great impact on various fields in physics \cite{Su:1979aa}. 
In particular, fractional fermion charges on solitons attracted considerable 
attention not only in condensed-matter physics 
\cite{Takayama:1980aa,Thouless:1983fk}  
but also in quantum field theory \cite{Goldstone:1981aa,niemisemenoff86R}.
Let us consider $(1+1)$-dimensional fermions coupled with two scaler fields \cite{Goldstone:1981aa}
\begin{alignat}1
\bar\psi(\varphi_1+\varphi_2i\gamma_5)\psi\propto \bar\psi e^{i\theta\gamma_5}\psi,
\label{DirMas}
\end{alignat}
where 
$\varphi_j$ and, hence, the parameter $\theta$ are assumed to be slowly varying in space and time.
Then, the expectation value of the current  $j^{\mu}\equiv \bar\psi\gamma^\mu\psi$ is given by \cite{Goldstone:1981aa}
\begin{alignat}1
\langle j^\mu(x,t)\rangle=\frac{1}{2\pi}\epsilon^{\mu\nu}\partial_\nu\theta.
\label{DirCur}
\end{alignat}
When $\theta$ is a function of $x$ with the boundary condition $\theta(+\infty)=\theta(-\infty)+\pi$, 
the fermion charge  $\nu$ becomes $\nu=-1/2$.
This  was discussed in terms of the fractional fermion charge in background soliton fields \cite{Goldstone:1981aa,niemisemenoff86R}.
Generically, Eq. (\ref{DirCur}) 
implies that when $\theta$ is a periodic function of $x$ with $\theta(x+L/2)=\theta(x-L/2)+2\pi p$, 
the fermion charge in one period becomes
\begin{alignat}1
\nu\equiv\int_{-L/2}^{L/2}\langle j^0(t,x)\rangle dx=-p.
\label{DirCha}
\end{alignat}  
The aforementioned fractional charge 
is regarded as the special case $L=\infty$ and $p=1/2$.
On the other hand, when $\theta$  is a function of $t$ with period $T$, 
$\theta(t+T)=\theta(t)\pm2\pi $, the pumped charge $c$ defined by
integrating $\langle j^1\rangle$ 
with respect to $t$ 
over one period becomes $c=\pm1$.
This is 
currently known as 
a Thouless pump \cite{Thouless:1983fk}.
Lattice Wilson--Dirac fermions 
indeed 
exhibit nontrivial Thouless pumps \cite{Fukui:2017aa}.
Thus, Eq. (\ref{DirCur}) 
implies that the fermion charge and the pumped charge 
have the same origin.

Generically, 
a Thouless pump is characterized by the 
Chern number of a gapped ground state, which is
the change of the 1D polarization as a function of $t$.
Bulk polarization 
is defined as the Berry phase \cite{Vanderbilt:1993fk,King-Smith:1993aa,Marzari:2012aa} at a fixed $t$.
For a system with boundaries, it corresponds to the center of mass (CM) \cite{Wang:2013fk_pump}, 
and its quantized shift was observed experimentally \cite{Nakajima:2016aa,Lohse:2016aa}.
Recently, quasi-periodic potentials have been introduced to study disorder effects 
in the Thouless pumping \cite{Marra:2020wh,Nakajima:2021xw}.

In this 
study, we generalize the RM model \cite{Rice:1982qf,Xiao:2010fk,Wang:2013fk_pump}
to include long-period spatial modulations 
and discuss the fermion charge with topological origin.
We show that the fermion charge and the pumped charge obey
a Diophantine equation similar to that for the Harper equation.
This leads to a one-dimensional analog of the Streda formula in the QHE \cite{Streda:1982aa}.
Namely, 
a change of the fermion charge is due to 
a change in
the periodicity of the spatial modulation.
For systems with charge conservation, 
a change of the fermion charge induces a change 
in the length of the fermion chain. 
We refer to these phenomena as spatial charge pumping, which provides a novel opportunity 
for experimental observation of topological pumping.  It also 
allows a direct measurement of the Streda formula in the QHE.

The RM model consists of alternating hoppings as well as staggered potentials.
Let $H(t)=\sum_{j} c_j^\dagger \hat{\cal H}_j(t) c_{j}$ be the Hamiltonian. 
Then, the Hamiltonian operator $\hat{\cal H}_j$ is written 
as 
\begin{alignat}1
\hat{\cal H}_j(t)=\frac{t_0+\cinb{\sigma^j}\delta_j(t)}{2}\hat\delta+\hat\delta^*\frac{t_0+\cinb{\sigma^j}\delta_j(t)}{2}
+\cinb{\sigma^j}\Delta_j(t),
\label{Ham}
\end{alignat}
where $\cinb{\sigma}=-1$, and $\hat\delta$ and $\hat\delta^*$ are forward and backward shift operators, respectively, 
acting on the right, that is, 
$\hat\delta f_j=f_{j+1}$ and $\hat\delta^* f_j=f_{j-1}$. 
The alternating hopping and staggered potential are defined by
\begin{alignat}1
&\delta_j(t)=\delta_0 \left[\cos 2\pi(t/T+p j/q)+r_h\right],
\nonumber\\
&\Delta_j(t)=\Delta_0\left[\sin 2\pi( t/T+p j/q)+r_m\right],
\label{PerPot}
\end{alignat}
where $q$ is assumed to be an even integer.
These are periodic functions of $t$ and $j$, as $\delta_j(t+T)=\delta_j(t)$ and $\delta_{j+q}(t)=\delta_j(t)$, 
and likewise for $\Delta_j(t)$.
This model will be referred to as the generalized RM (gRM) model. 
When $p=0$, it reduces to the conventional RM model.

First, we show the relationship between the fermion charge $\nu$ and the pumped charge $c$.
We temporarily assume that $q$ is sufficiently large compared with $p$.
Then, the Hamiltonian depends on $j$ smoothly through $(p/q)j$ but rapidly through $(-)^j$ .
To describe the latter,  we introduce a 2-unit cell including two neighboring 
sites $(2j-1,2j)$
and the corresponding local charge operator $\rho_{j}\equiv c_{2j-1}^\dagger c_{2j-1}+c_{2j}^\dagger c_{2j}$.
From the Heisenberg equation $i\hbar\partial_t\rho_j=[H(t),\rho_j]$, the continuity equation reads
\begin{alignat}1
\partial_t\rho_{j}+ (\cinb{I_{j}-I_{j-1}})=0,
\label{ConEqu}
\end{alignat}
where the local current operator is defined by $\cinb{I_{j}}=(i/\hbar)t_{2j}(c_{2j+1}^\dagger c_{2j}-\mbox{H.c})$
with $t_j=\left(t_0+\sigma^j\delta_j\right)/2$.
We note that $t_{2j}$ 
depends on $\xi\equiv t/T+2pj/q$ smoothly, 
so that the current operator is denoted as $\cinb{I_j}(\xi)$.
For sufficiently large $q$, we have $t_{2j-2}\sim t_{2j}-(2p/q)\partial_\xi t_{2j}$. 
Thus, Eq. (\ref{ConEqu}) becomes
\begin{alignat}1
\partial_t\rho_{j}+ \cinb{I_{j}(\xi)-I_{j-1}(\xi)}+(2p/q)\partial_\xi \cinb{I_{j-1}}(\xi)=0,
\end{alignat}
where 
$\xi$ is regarded as a fixed parameter independent of $j$.
We 
define averaged expectation values $\sum_j\langle\rho_{j}\rangle /(N_{\rm s}/2)\equiv \rho(\xi)$ and, 
$\sum_j\langle \cinb{I_{j}}\rangle /(N_{\rm s}/2)\equiv \cinb{I(\xi)}$, where 
$j$ runs
over all $N_{\rm s}/2$ unit cells along the chain. Then, we have
\begin{alignat}1
\partial_t\rho(\xi)+ (2p/q)\partial_\xi \cinb{I(\xi)}=0.
\end{alignat}
It follows from $\partial_t=\partial_\xi/T$ that 
\begin{alignat}1
\rho(\xi)/T+(2p/q)\cinb{I(\xi)}=\mbox{const.}
\label{Con}
\end{alignat}
For a large $q$, let us define the fermion charge in the $q$-unit cell, $\nu$, which is referred to as a filling number: 
\begin{alignat}1
\nu&=\sum_{j=1}^{q/2}\rho(\xi\cinn{=t/T+2pj/q})=\frac{q}{2p}\int_0^{p}d\xi\rho(\xi)
\nonumber\\
&=-T\int_0^{p}d\xi \cinb{I(\xi)}=-\int_0^{pT}\cinb{I(\xi)}dt=-pc,
\label{q0c}
\end{alignat}
where 
\cinn{we have neglected the constant term in Eq. (\ref{Con}), and} $c$ 
denotes the pumped charge, which is in fact 
the Chern number \cite{Thouless:1983fk}. 

\begin{figure}[h]
\begin{center}
\begin{tabular}{cc}
\begin{minipage}{0.5\linewidth}
\includegraphics[width=\linewidth]{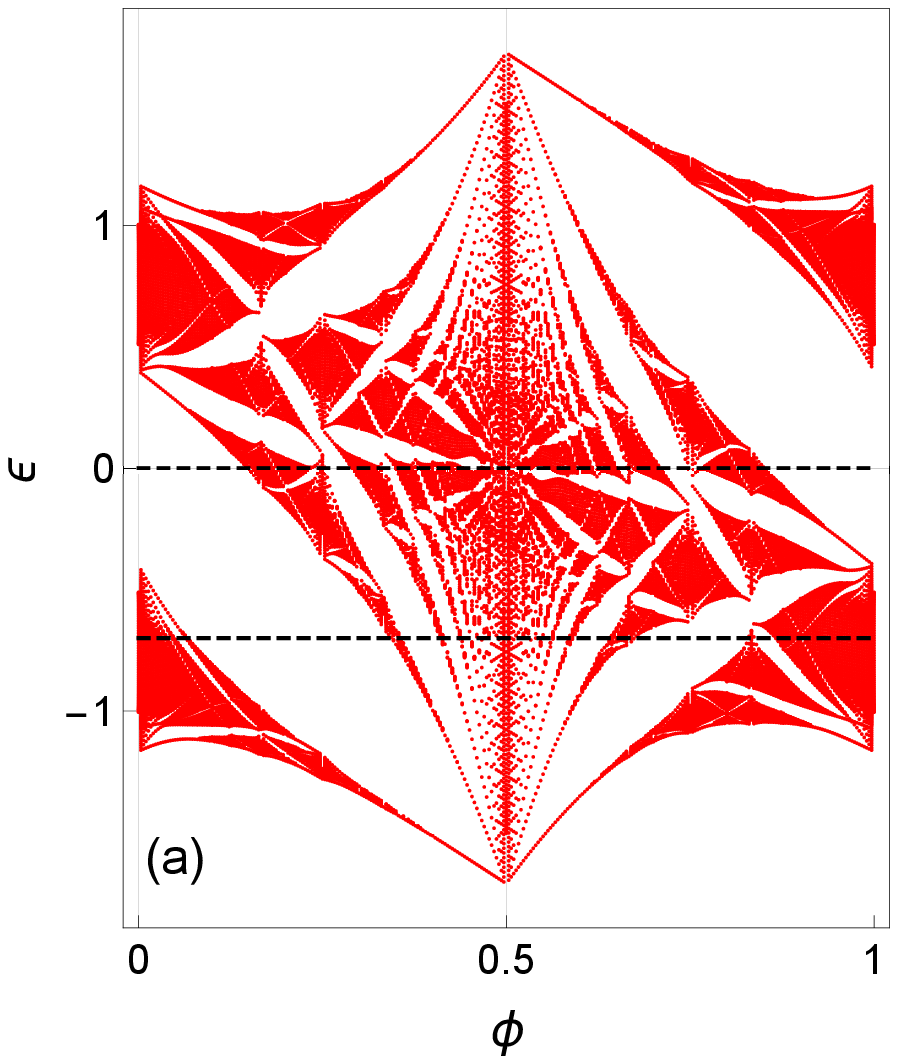}
\end{minipage}
&
\begin{minipage}{0.5\linewidth}
\begin{tabular}{c}
\includegraphics[width=\linewidth]{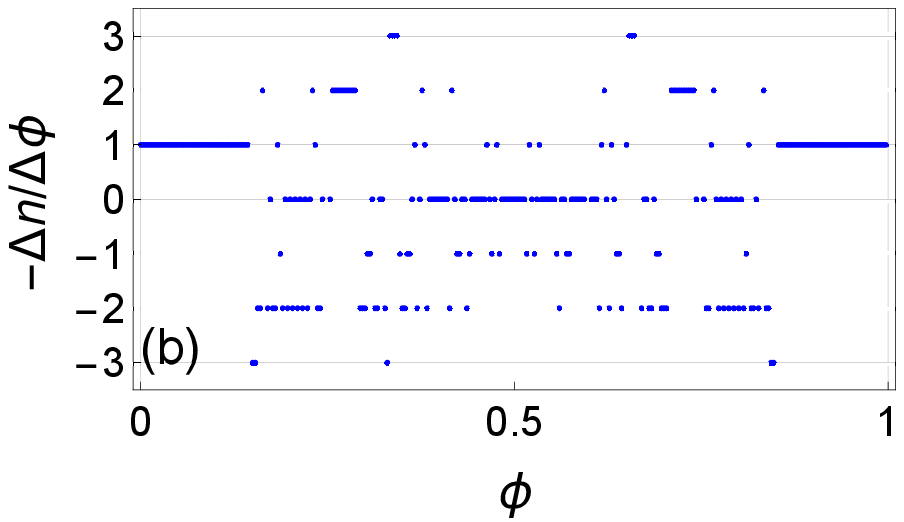}\\
\includegraphics[width=\linewidth]{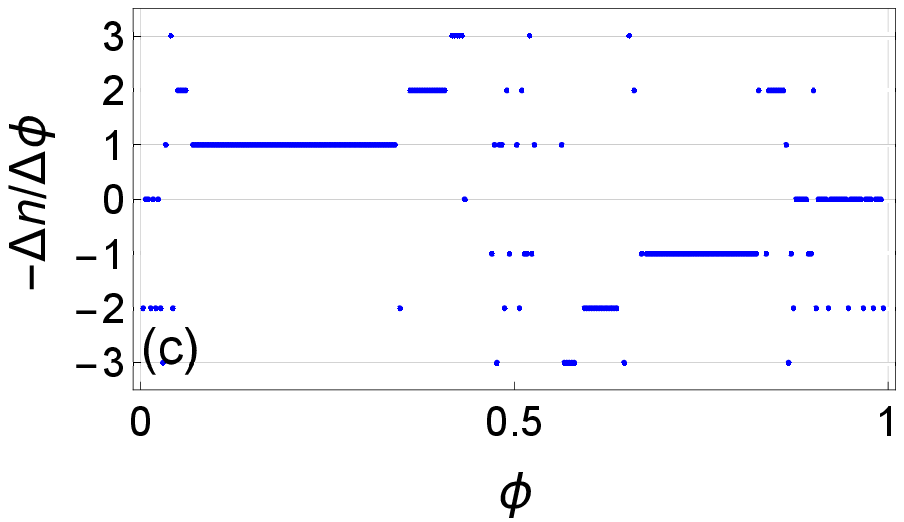}\\
\end{tabular}
\end{minipage}
\end{tabular}
\caption{
(a) Spectrum at $t=0$ as a function of $\phi=p/q$. The dashed lines denote the Fermi energies
along which the Chern numbers in (b) and (c) are computed.
(b) and (c) show the Chern numbers as a function of $\phi$ at fixed Fermi energies
$0$ and $-0.7$, respectively.
The parameters used are $t_0=1$, $\delta_0=\Delta_0=0.5$, $r_h=0$, and $r_m=0.2$.
}
\label{f:but}
\end{center}
\end{figure}

Equation (\ref{q0c})  provides a method of computing the Chern number from the filling number.
To this end, it is convenient to define the fermion charge density
$n\equiv \nu/q$ at a given time, for instance,
$t=0$. Then, Eq. (\ref{q0c}) reads
\begin{alignat}1
\frac{\partial n}{\partial \phi}=-c,\quad \phi=p/q.
\label{Str}
\end{alignat}
This serves as the Streda formula in the QHE \cite{Streda:1982aa}.
It should be noted that $\nu$ and, hence, $n$ are given at a fixed time. 
Therefore, one can compute the Chern numbers without using any time dependence.
In Fig. \ref{f:but} (a), the spectrum at $t=0$, and in (b) and (c),
the Chern numbers computed by Eq. (\ref{Str}) at fixed Fermi energies, are shown as functions of $\phi$.
It can be seen that when the Fermi energy lies within a gap, the left-hand side of Eq. (\ref{Str})
is stably constant, from which 
the Chern number of the gap can be obtained.
The Streda formula (\ref{Str}) is in sharp contrast to the conventional method of computing Chern numbers 
using Bloch wave functions in the $k$-$t$ space \cite{Thouless:1983fk}.

Interestingly, Eq. (\ref{Str}) provides an alternative opportunity to observe 
topological pumps 
by controlling $\phi$  in Eq. (\ref{PerPot}) 
with $t$ fixed. 
\cin{We 
assume that there is a chain with $N$ fermions confined in a harmonic trap. 
We also assume that the harmonic trap is so weak that all fermions occupy the bands up to the $\nu$th, and 
there is a large gap above the $\nu$th band.
Then, the length of the chain is $L/a=N/n$, where $a$ is the lattice constant, and $n=\nu/q$ is the charge density.}
Now, we consider an adiabatic insertion of $\Delta\phi$ into this system through Eq. (\ref{PerPot}). 
\cin{This induces 
a charge-density change $n\rightarrow n'=n+\Delta n$ 
by Eq. (\ref{PerPot}), 
where $\Delta n=-c\Delta\phi$.}
\cin{
Owing to the charge conservation of the present system, 
the charge-density change implies 
a length change 
$L\rightarrow L'=L+\Delta L$, and hence,}
we have $nL/a\cin{=n'L'/a}=(n+\Delta n)(L+\Delta L)/a$,
which yields
\begin{alignat}1
\Delta L/L=(c/n)\Delta\phi.
\label{SpaPum}
\end{alignat}
This equation 
implies that the length of the fermion chain becomes longer by the amount $\Delta L$ 
because of the adiabatic insertion of $\Delta\phi$. 
This is one of our main results.
\cin{Let us apply this 
to the conventional RM model.}
When $\phi=0$, the RM model 
yields $c=1$ at $n=1/2$.
\cin{As above,
we assume a chain of $N$ fermions of length $L/a=2N$ in a harmonic trap. 
The gap is also assumed to be so large that all fermions occupy the lower band. 
This was indeed realized in the experiment \cite{Nakajima:2016aa}.
We 
note that in the case of $\phi=0$, we can 
choose any $q$; 
thus, let us set $q=L/a$.}
Then, an adiabatic $\Delta\phi=p/q$ insertion yields $\Delta L/a=2p$, 
implying that the length of the  chain becomes longer by $2p$.
\cin{For a more generic choice of $q$, see below Eq. (\ref{Dio}).}
We can interpret this as a spatial charge pump: 
The charge at the right end is pumped $2p$ to the right compared with the charge at the left end.
This causes the change in the length of the system.
On the other hand, in 
the Thouless pump, each charge is pumped in the same direction by the same amount under time evolution, 
so that the length of the system 
does not change, but the CM is shifted instead. 
In this sense, we refer to the Thouless pump as a temporal pump to distinguish it 
from the aforementioned spatial pump
owing to the adiabatic change of $\phi$.

\begin{figure}[h]
\begin{center}
\begin{tabular}{cc}
\includegraphics[width=0.5\linewidth]{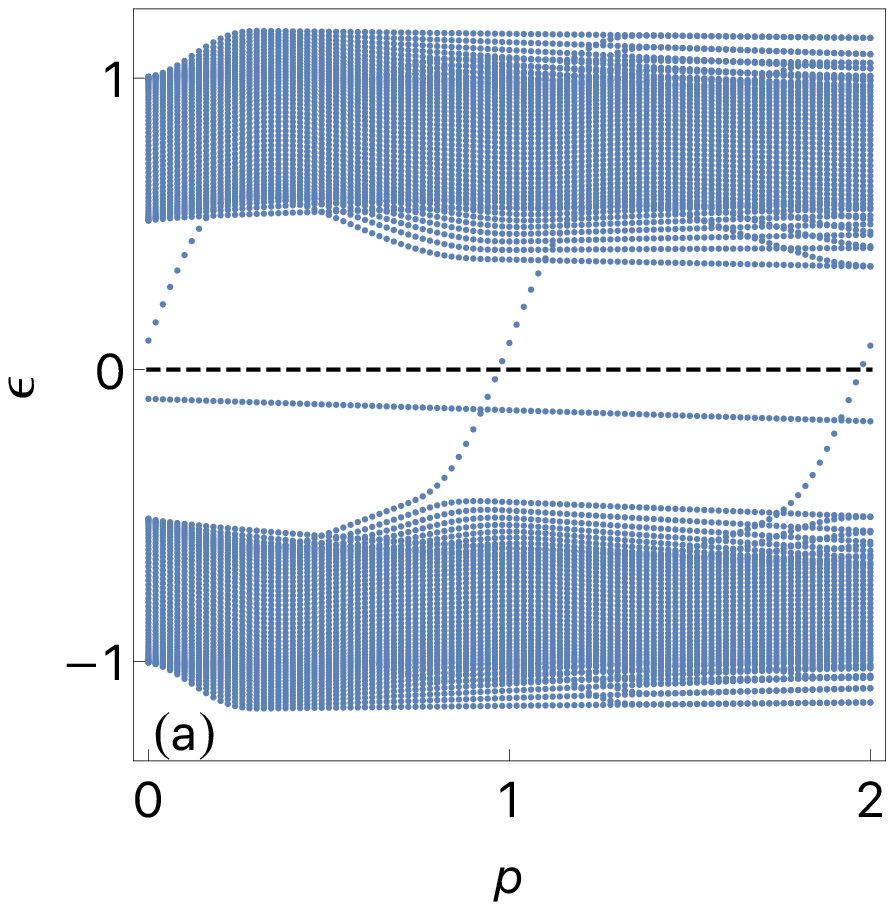}
&
\includegraphics[width=0.5\linewidth]{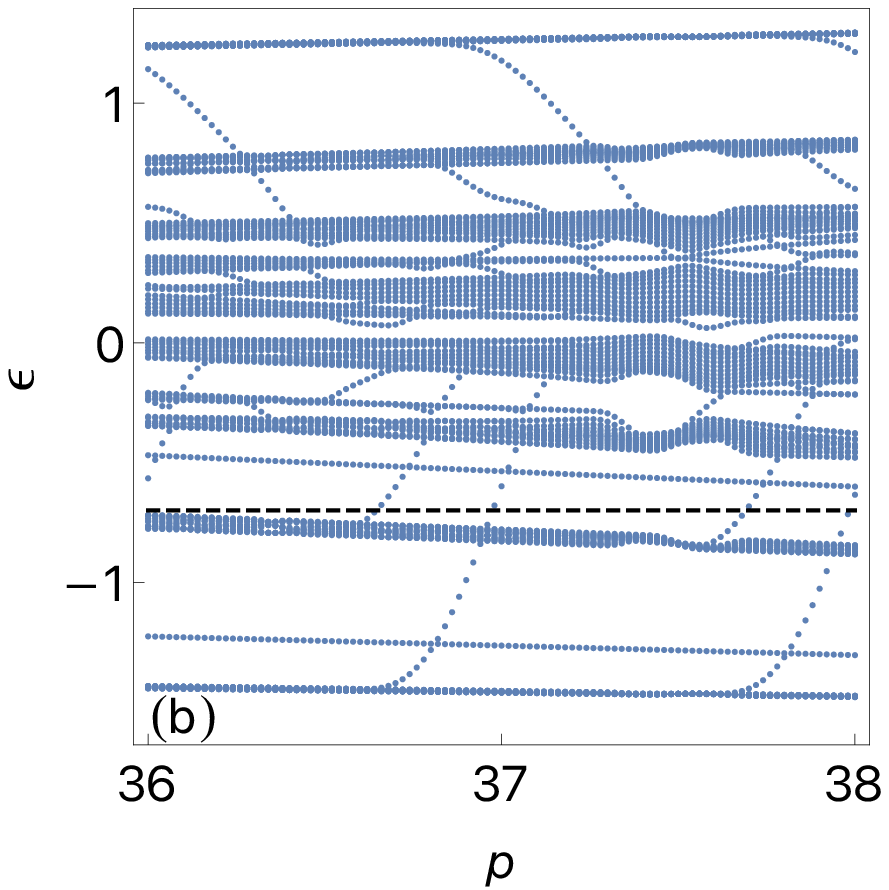}
\\
\includegraphics[width=0.5\linewidth]{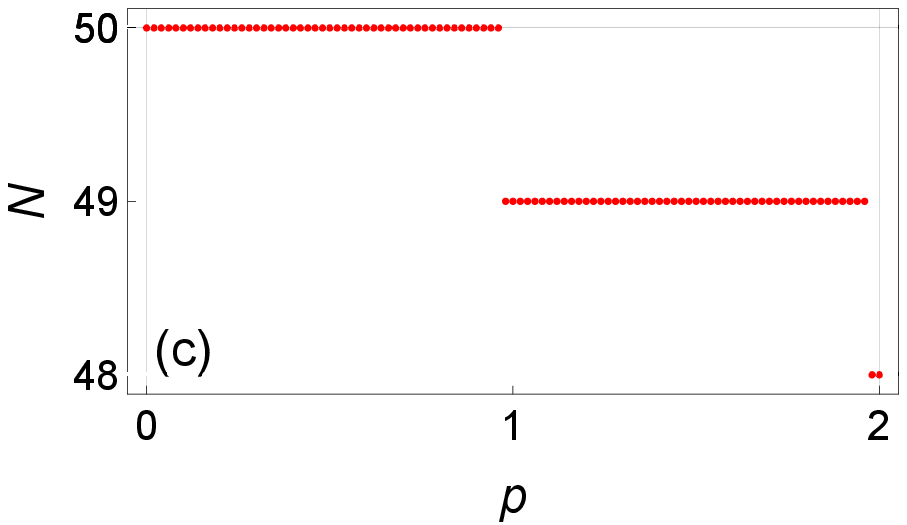}
&
\includegraphics[width=0.5\linewidth]{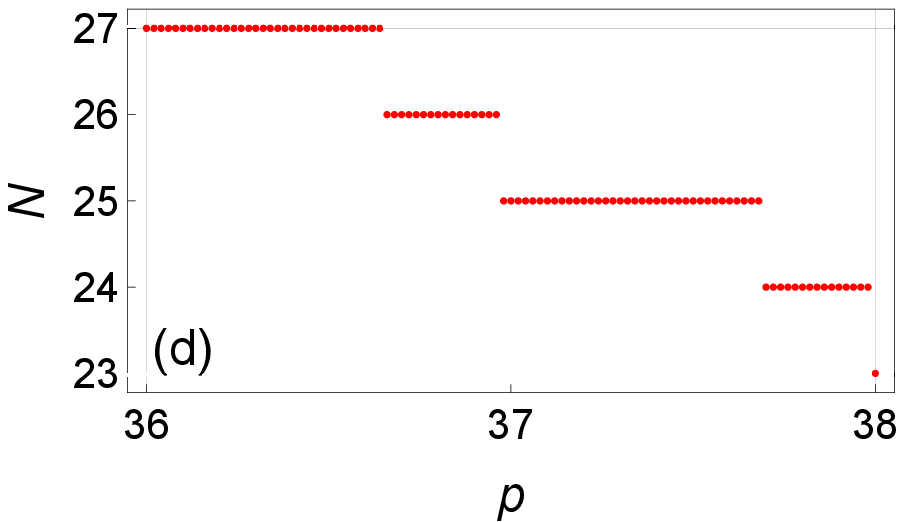}
\\
\includegraphics[width=0.5\linewidth]{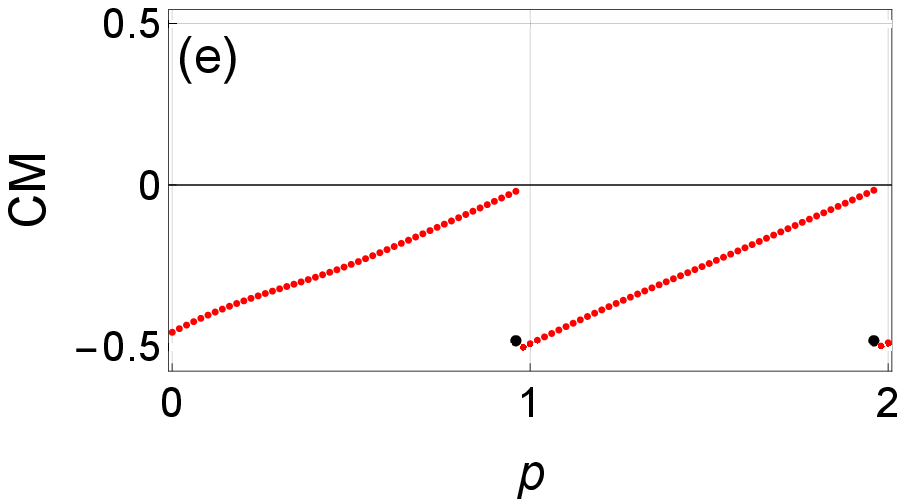}
&
\includegraphics[width=0.5\linewidth]{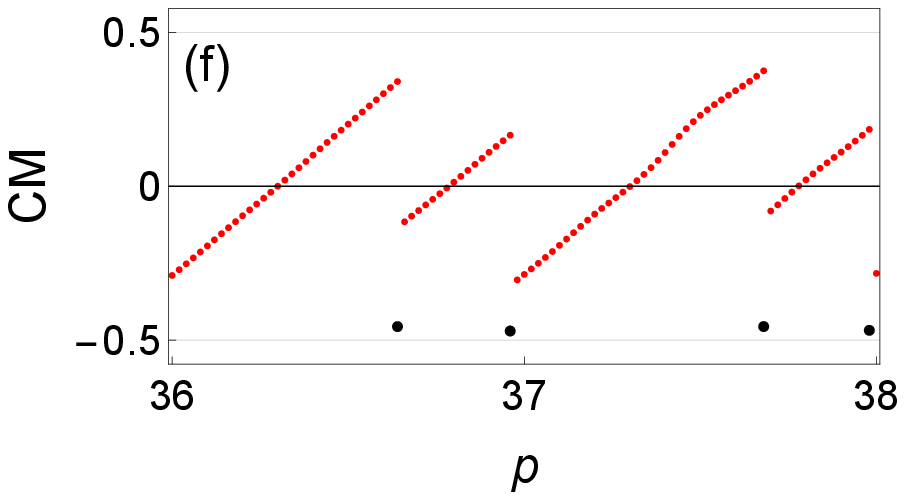}
\end{tabular}
\caption{
(a) and (b) show energy spectra as functions of $p$  with $q=100$ fixed. 
The length $L/a$ of the system is 
chosen  as $L/a=q$, and open boundary conditions are imposed on both ends.
Dashed lines denote the same Fermi energies as in Fig. \ref{f:but}, which are located in gaps with $c=1$ and $2$;
(c) and (d) show the number of states below the Fermi energies specified in (a) and (b), respectively;
(e) and (f) show the CM of the occupied states in (a) and (b), respectively.
The black dots
indicate the values of the jumps at the discontinuity of the CM.
}
\label{f:edge}
\end{center}
\end{figure}

We can also consider the spatial pump from the 
perspective of the bulk-edge correspondence.
Specifically, we 
consider the same system mentioned above, 
that is, 
the system of density $n$ and size $L$ 
that includes $N=nL/a$ fermions, on which the open boundary conditions are imposed. 
Now, we assume that the system is in contact with a particle reservoir.
It follows from Eq. (\ref{Str}) that 
the adiabatic $\Delta\phi=\Delta p/q$ ($q=L/a$) insertion causes 
a change of the total fermion charge,
\begin{alignat}1
\Delta N=-c\Delta p.
\end{alignat}
Such a change  is due to 
the spectral flow of the edge states across the Fermi energy when $p$ is varied.
In Figs. \ref{f:edge} (a) and (b), the spectra of the systems of size $L/a=q=100$ as functions of $p$ \cin{are shown}.
It is clearly seen 
that 
the number of states below the Fermi energy only decreases 
by $c$ for every $\Delta p=1$ 
because of the
edge states across the Fermi energy. 
Furthermore, 
these 
edge states are localized at the right end, 
as the CM
exhibits discontinuous changes by $-1/2$  \cite{Hatsugai:2016aa}, as shown 
in Figs. \ref{f:edge} (e) and (f).
Here, the 
CM is defined to be $-1/2$ at $j=1$, and $+1/2$ at $j=L/a$.

So far, we have discussed the spatial pump based on 
the filling number (\ref{q0c}) and the resultant Streda formula (\ref{Str}).
To apply Eq. (\ref{q0c}) to the lattice model more precisely, some corrections and definitions are 
required. 
First,  when $p=0$, $\nu$ is the number of occupied bands for the RM model;
 completely unoccupied (0), half-filled (1), and fully occupied (2). 
In the gRM model in Eq. (\ref{PerPot}), these filling numbers are translated into $0$, $q/2$, and $q$. 
This should be included in Eq. (\ref{q0c}), probably as the constant term in Eq. (\ref{Con}) neglected in Eq. (\ref{q0c}).
Second, the gRM model 
has $q$ bands generically, so that Eq. (\ref{q0c}) is expected to hold at each gap separately.
Let us  
focus on 
the $i$th  gap.
The filling number $\nu_i$ is the number of bands below the $i$th gap, and the pumped charge $c_i$
is the Chern number of the gap, 
that is, the sum of the Chern numbers of all bands below the $i$th gap.
With these corrections and definitions, we propose the following relationship between $\nu_i$ and $c_i$:
\begin{alignat}1
\nu_i=s_iq/2-pc_i,
\label{Dio}
\end{alignat}
where $s_i$ 
denotes a certain integer. 
This equation is a Diophantine equation similar to that \cite{Thouless:1982uq} for the Harper equation \cite{Harper:1955aa}.
Using this, Eq. (\ref{SpaPum}) 
provides a more precise length  of the system \cin{as follows:
Let us 
begin with a model including $\phi_0$, and control the fermion number $N$
such that 
all fermions occupy  the bands below the $i$th gap, implying the charge density $n_i=s_i/2-\phi_0c_i$. 
This may be possible when the $i$th gap is sufficiently large.
Then, the initial length of the chain is $L_0/a=N/n_i$.
We 
assume that we change $\phi_0\rightarrow \phi$ and 
obtain $n_{i'}=s_{i'}/2-\phi c_{i'}$.  The length becomes $L/a=N/n_{i'}$. 
It then follows that 
 }
\begin{alignat}1
L/L_0=(s_i-2c_i\phi_0)/(s_i-2c_i\phi),
\end{alignat}
\cin{provided that $c_{i'}=c_i$ and $s_{i'}=s_i$, implying that all fermions continue to occupy the bands up to the same topological gap
characterized by $c_i$ \cinn{and $s_i$}. 
For the conventional half-filled 
RM model discussed below Eq. (\ref{SpaPum}), we have $L/L_0=1/(1-2\phi)$, where $L_0$ is the length of the system when $\phi_0=0$.
For a small $\phi$, this relation is approximated by  $L=2 \phi L_0 $, and if we 
choose $q$ 
in $\phi=p/q$ as $q=N/2 (=L_0/a)$, 
the previous result can be reproduced.
}

\begin{figure}[h]
\begin{center}
\begin{tabular}{cc}
\includegraphics[width=0.5\linewidth]{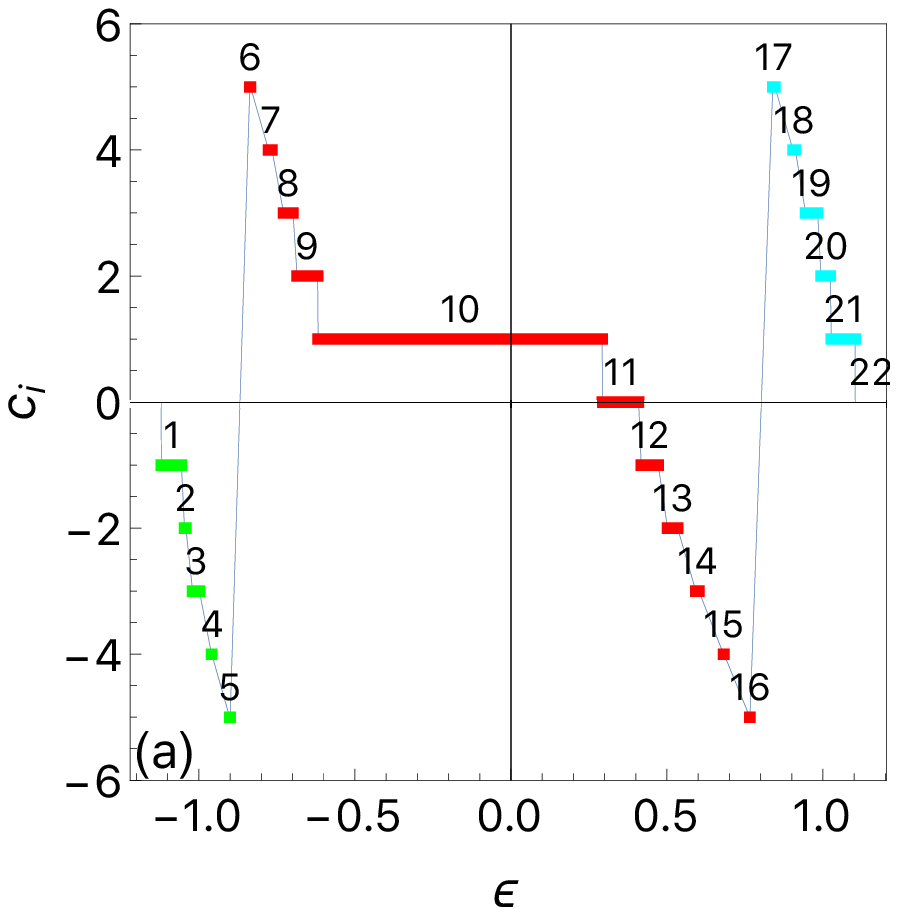}
&
\includegraphics[width=0.5\linewidth]{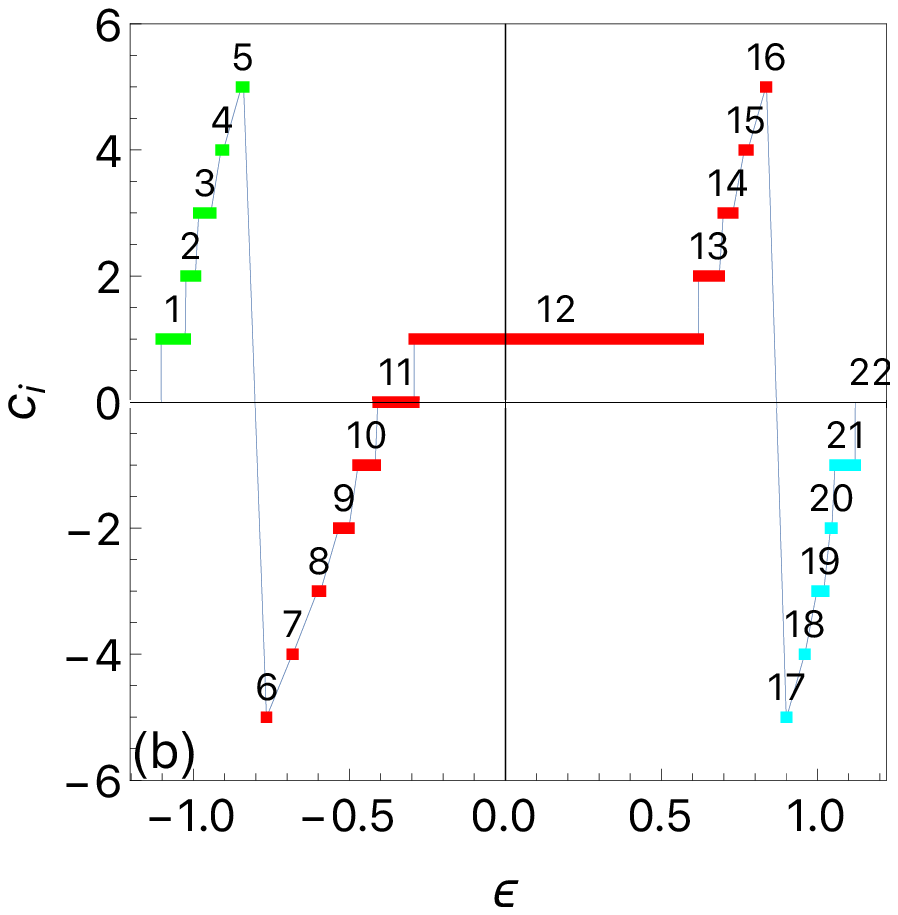}
\end{tabular}
\caption{
Chern numbers \cinb{of gaps} as functions of the Fermi energy $\epsilon$ for the model with $q=22$ and (a) $p=1$, (b) $p=-1$.
Horizontal colored thick lines 
indicate gaps, and the number above the $i$th gap 
is $\nu_i$.
The green, red, and cyan 
lines 
correspond to the gaps satisfying Eq. (\ref{Dio}) with $s_i=0,1$, and $2$, respectively.
The parameters of the model are same as those in Fig. \ref{f:but}.
}
\label{f:ch_1}
\end{center}
\end{figure}

To 
verify the validity of the formula (\ref{Dio}), 
the filling number $\nu_i$ as well as the Chern numbers $c_i$ 
\cite{Thouless:1983fk} 
are shown in Fig. \ref{f:ch_1} in the case  of $p/q=\pm1/22$.
Here, the Chern numbers 
are those defined by the Berry curvature of the Bloch states in the $k$-$t$ space 
\cite{Thouless:1982uq,Thouless:1983fk}, calculated using the method in \cite{FHS05}.
It can be seen that the system has $q$ separated bands: 
Each band has 
Chern number $\pm1$, 
and as a result, the Chern number of the gap as a function of energy 
exhibits a step-like behavior 
similar to that of the Landau levels in the QHE. 
Such a step-like structure is divided into three regions by the van Hove singularities, which carry 
large Chern numbers \cite{Hatsugai:2006aa}.
Each region is characterized by $s_i$: 
$s_i=0$  $(2)$ series is associated with particle (hole) states 
in the fully-unoccupied (fully-occupied) vacuum, 
whereas $s_j=1$ series is associated with the half-filled ground state of the RM model. 
\begin{figure}[h]
\begin{center}
\begin{tabular}{cc}
\includegraphics[width=0.5\linewidth]{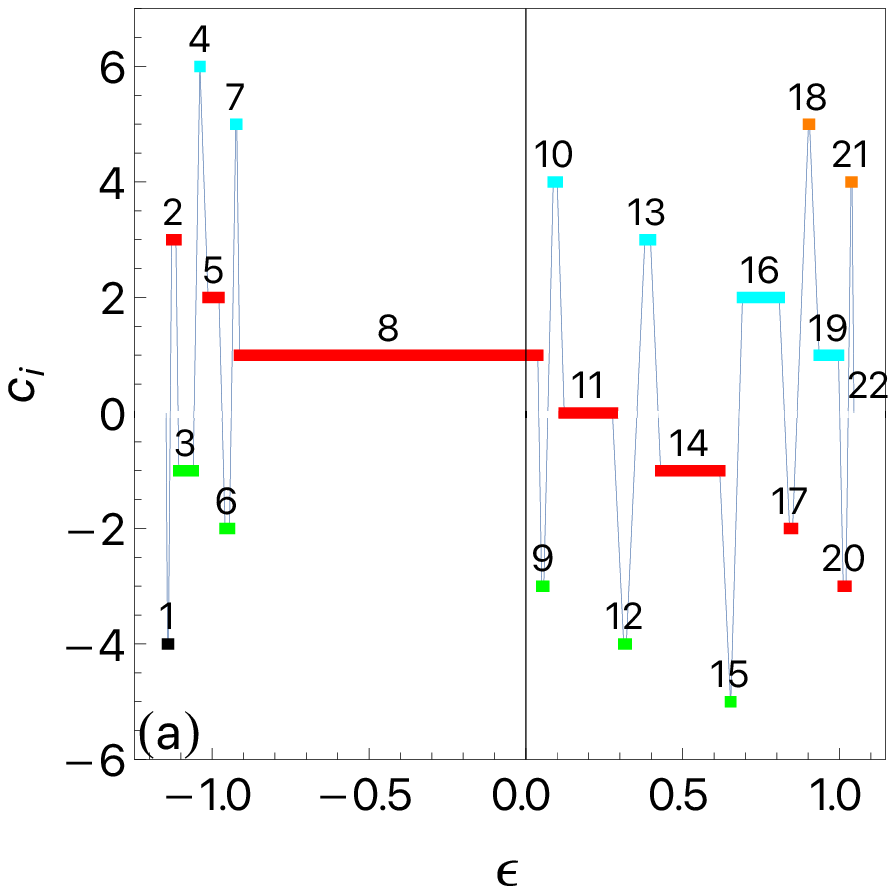}
&
\includegraphics[width=0.5\linewidth]{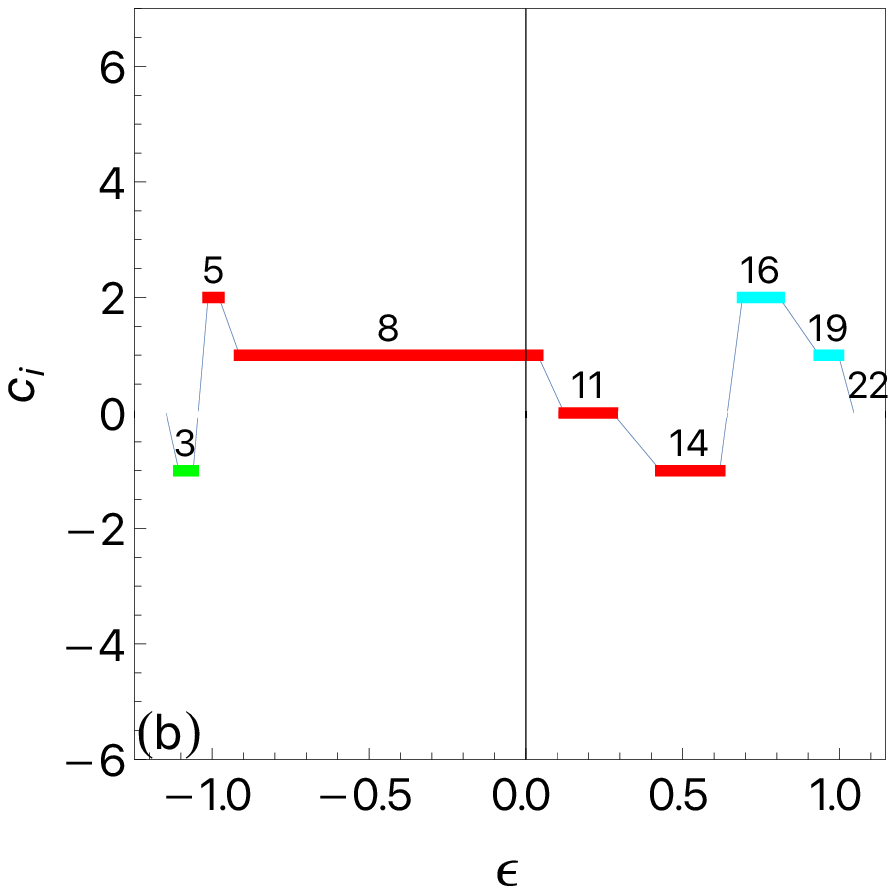}
\end{tabular}
\caption{
Same as in Fig. \ref{f:ch_1} but for $p=3$. In (a), the orange \cinb{lines} \cinb{and the black line} 
correspond to $s_i=3$ \cinb{and $s_j=-1$}, respectively. Other colors 
have the same interpretation as 
in Fig. \ref{f:ch_1}.
\cinb{
It can be seen  that the Chern number of  each band, calculated by the difference of the gap Chern numbers 
$c_i-c_{i-1}$, is $-4, 7, -4, 7, -4, -4, \cdots$, starting from the 1st band;}
(b) is the same system as (a) but 
uses a larger threshold gap: Here, the threshold gap 
implies that  
any two bands are regarded as degenerate 
if the minimum direct gaps between them are smaller than the threshold gap. 
}
\label{f:ch_3}
\end{center}
\end{figure}
Other examples are shown in Fig. \ref{f:ch_3}. When $p=3$, each band has large Chern numbers, namely, 7 or $-4$,
as seen in Fig. \ref{f:ch_3} (a), and gaps with various $s_i$ coexist in the same energy region.
Nevertheless, if
tiny gaps are ignored and 
regarded as degenerate bands,
as 
shown in Fig. \ref{f:ch_3} (b), 
three series of gaps separated by van Hove singularities can be seen, 
as in the case of $p=1$ in Fig. \ref{f:ch_1}.
These examples as well as various other numerical studies 
involving small $q$ and/or large $p$ 
demonstrate that the Diophantine equation (\ref{Dio}) is valid generically for the gRM model.

The effect of the spatial pump can be 
understood as 
a change of the filling number $\nu_i$ for nonzero $\phi$:
In Figs. \ref{f:ch_1} and \ref{f:ch_3}, 
\cinb{
it can be seen that the the gaps at zero energy have filling numbers $\nu=10$, $12$, and $8$ 
for $p=1$, $-1$, and $3$, respectively, as well as the same Chern number $c=1$. 
This gap corresponds to the half-filled gap with $\nu=11$, when $\phi=0$ ($p=0$ and $q=22$).
Therefore, in all these cases, we have
the filling-number shift 
$\Delta\nu=\pm1, -3=-c\Delta p$. 
} 
This is the effect of the spatial pump from the bulk 
perspective.

As is the case with the RM model \cite{Wang:2013fk_pump}, the present system 
would be realized in cold atoms using the optical superlattice \cite{Nakajima:2016aa,Lohse:2016aa}
\begin{alignat}1
V(x,\phi,t)=&-v_1 \cos^2\frac{\pi x}{a}-v_2 \cos^2\left[\left(\frac{1}{2}+\phi\right)\frac{\pi x}{a}+\frac{\pi t}{T}\right]
\nonumber\\
&\cin{-v_3\cos^2\left(\frac{\pi x}{2a}+\theta\right),}
\end{alignat} 
\cin{where $\theta$ controls the parameters $r_h$ and $r_m$ in Eq. (\ref{PerPot}).}
An adiabatic change of $t$ with $\phi=0$ fixed
yields the temporal Thouless pump already observed \cite{Nakajima:2016aa,Lohse:2016aa}, whereas 
an adiabatic change of  $\phi$ with $t$ fixed would yield the spatial pump proposed in this paper.
\cinn{Here,  $\phi$ could be controlled by the change in the tilted angle between the optical superlattices \cite{Marra:2020wh,Nakajima:2021xw}.}
In particular, the latter could provide a direct measurement of the Streda formula in 1D systems.

In conclusion, we 
discussed the topological aspect of  the filling number or fermion charge of the gRM model.
We 
derived 
a Diophantine equation 
relating the filling number and the pumped charge, and the resultant 
Streda formula.
These predict a spatial pump, which could be observed as a change in
the length of the fermion chain.
A simpler model can be considered if 
$\cinb{\sigma=1}$ is chosen in Eq. (\ref{Ham}). 
This model is quite analogous to the conventional QHE, 
as Eq.  (\ref{Dio})  is modified into
$\nu_i=s_iq-pc_i$,
which is the same as the Diophantine equation 
for the Harper equation.
Such systems with $\cinb{\sigma=\pm1}$ would offer a promising platform for studying more intimate relationships between
topological pumps and the QHE.

TF would like to thank Y. Takahashi for fruitful discussions.
This work was supported in part by Grants-in-Aid for Scientific Research Numbers 17H06138
from Japan Society for the Promotion of Science.


\begin{thebibliography}{22}
\expandafter\ifx\csname natexlab\endcsname\relax\def\natexlab#1{#1}\fi
\expandafter\ifx\csname bibnamefont\endcsname\relax
  \def\bibnamefont#1{#1}\fi
\expandafter\ifx\csname bibfnamefont\endcsname\relax
  \def\bibfnamefont#1{#1}\fi
\expandafter\ifx\csname citenamefont\endcsname\relax
  \def\citenamefont#1{#1}\fi
\expandafter\ifx\csname url\endcsname\relax
  \def\url#1{\texttt{#1}}\fi
\expandafter\ifx\csname urlprefix\endcsname\relax\def\urlprefix{URL }\fi
\providecommand{\bibinfo}[2]{#2}
\providecommand{\eprint}[2][]{\url{#2}}

\bibitem[{\citenamefont{Su et~al.}(1979)\citenamefont{Su, Schrieffer, and
  Heeger}}]{Su:1979aa}
\bibinfo{author}{\bibfnamefont{W.~P.} \bibnamefont{Su}},
  \bibinfo{author}{\bibfnamefont{J.~R.} \bibnamefont{Schrieffer}},
  \bibnamefont{and} \bibinfo{author}{\bibfnamefont{A.~J.}
  \bibnamefont{Heeger}}, \bibinfo{journal}{Physical Review Letters}
  \textbf{\bibinfo{volume}{42}}, \bibinfo{pages}{1698} (\bibinfo{year}{1979}).

\bibitem[{\citenamefont{Takayama et~al.}(1980)\citenamefont{Takayama, Lin-Liu,
  and Maki}}]{Takayama:1980aa}
\bibinfo{author}{\bibfnamefont{H.}~\bibnamefont{Takayama}},
  \bibinfo{author}{\bibfnamefont{Y.~R.} \bibnamefont{Lin-Liu}},
  \bibnamefont{and} \bibinfo{author}{\bibfnamefont{K.}~\bibnamefont{Maki}},
  \bibinfo{journal}{Physical Review B} \textbf{\bibinfo{volume}{21}},
  \bibinfo{pages}{2388} (\bibinfo{year}{1980}).

\bibitem[{\citenamefont{Thouless}(1983)}]{Thouless:1983fk}
\bibinfo{author}{\bibfnamefont{D.~J.} \bibnamefont{Thouless}},
  \bibinfo{journal}{Physical Review B} \textbf{\bibinfo{volume}{27}},
  \bibinfo{pages}{6083} (\bibinfo{year}{1983}).

\bibitem[{\citenamefont{Goldstone and Wilczek}(1981)}]{Goldstone:1981aa}
\bibinfo{author}{\bibfnamefont{J.}~\bibnamefont{Goldstone}} \bibnamefont{and}
  \bibinfo{author}{\bibfnamefont{F.}~\bibnamefont{Wilczek}},
  \bibinfo{journal}{Physical Review Letters} \textbf{\bibinfo{volume}{47}},
  \bibinfo{pages}{986} (\bibinfo{year}{1981}).

\bibitem[{\citenamefont{Niemi and Semenoff}(1986)}]{niemisemenoff86R}
\bibinfo{author}{\bibfnamefont{A.~J.} \bibnamefont{Niemi}} \bibnamefont{and}
  \bibinfo{author}{\bibfnamefont{G.~W.} \bibnamefont{Semenoff}},
  \bibinfo{journal}{Phys. Rep.} \textbf{\bibinfo{volume}{135}},
  \bibinfo{pages}{99} (\bibinfo{year}{1986}).

\bibitem[{\citenamefont{Fukui and Fujiwara}(2017)}]{Fukui:2017aa}
\bibinfo{author}{\bibfnamefont{T.}~\bibnamefont{Fukui}} \bibnamefont{and}
  \bibinfo{author}{\bibfnamefont{T.}~\bibnamefont{Fujiwara}},
  \bibinfo{journal}{Physical Review B} \textbf{\bibinfo{volume}{96}},
  \bibinfo{pages}{205404} (\bibinfo{year}{2017}).

\bibitem[{\citenamefont{Vanderbilt and King-Smith}(1993)}]{Vanderbilt:1993fk}
\bibinfo{author}{\bibfnamefont{D.}~\bibnamefont{Vanderbilt}} \bibnamefont{and}
  \bibinfo{author}{\bibfnamefont{R.~D.} \bibnamefont{King-Smith}},
  \bibinfo{journal}{Physical Review B} \textbf{\bibinfo{volume}{48}},
  \bibinfo{pages}{4442} (\bibinfo{year}{1993}).

\bibitem[{\citenamefont{King-Smith and Vanderbilt}(1993)}]{King-Smith:1993aa}
\bibinfo{author}{\bibfnamefont{R.~D.} \bibnamefont{King-Smith}}
  \bibnamefont{and}
  \bibinfo{author}{\bibfnamefont{D.}~\bibnamefont{Vanderbilt}},
  \bibinfo{journal}{Physical Review B} \textbf{\bibinfo{volume}{47}},
  \bibinfo{pages}{1651} (\bibinfo{year}{1993}).

\bibitem[{\citenamefont{Marzari et~al.}(2012)\citenamefont{Marzari, Mostofi,
  Yates, Souza, and Vanderbilt}}]{Marzari:2012aa}
\bibinfo{author}{\bibfnamefont{N.}~\bibnamefont{Marzari}},
  \bibinfo{author}{\bibfnamefont{A.~A.} \bibnamefont{Mostofi}},
  \bibinfo{author}{\bibfnamefont{J.~R.} \bibnamefont{Yates}},
  \bibinfo{author}{\bibfnamefont{I.}~\bibnamefont{Souza}}, \bibnamefont{and}
  \bibinfo{author}{\bibfnamefont{D.}~\bibnamefont{Vanderbilt}},
  \bibinfo{journal}{Reviews of Modern Physics} \textbf{\bibinfo{volume}{84}},
  \bibinfo{pages}{1419} (\bibinfo{year}{2012}).

\bibitem[{\citenamefont{Wang et~al.}(2013)\citenamefont{Wang, Troyer, and
  Dai}}]{Wang:2013fk_pump}
\bibinfo{author}{\bibfnamefont{L.}~\bibnamefont{Wang}},
  \bibinfo{author}{\bibfnamefont{M.}~\bibnamefont{Troyer}}, \bibnamefont{and}
  \bibinfo{author}{\bibfnamefont{X.}~\bibnamefont{Dai}},
  \bibinfo{journal}{Physical Review Letters} \textbf{\bibinfo{volume}{111}},
  \bibinfo{pages}{026802} (\bibinfo{year}{2013}).

\bibitem[{\citenamefont{Nakajima et~al.}(2016)\citenamefont{Nakajima, Tomita,
  Taie, Ichinose, Ozawa, Wang, Troyer, and Takahashi}}]{Nakajima:2016aa}
\bibinfo{author}{\bibfnamefont{S.}~\bibnamefont{Nakajima}},
  \bibinfo{author}{\bibfnamefont{T.}~\bibnamefont{Tomita}},
  \bibinfo{author}{\bibfnamefont{S.}~\bibnamefont{Taie}},
  \bibinfo{author}{\bibfnamefont{T.}~\bibnamefont{Ichinose}},
  \bibinfo{author}{\bibfnamefont{H.}~\bibnamefont{Ozawa}},
  \bibinfo{author}{\bibfnamefont{L.}~\bibnamefont{Wang}},
  \bibinfo{author}{\bibfnamefont{M.}~\bibnamefont{Troyer}}, \bibnamefont{and}
  \bibinfo{author}{\bibfnamefont{Y.}~\bibnamefont{Takahashi}},
  \bibinfo{journal}{Nat Phys} \textbf{\bibinfo{volume}{12}},
  \bibinfo{pages}{296} (\bibinfo{year}{2016}).

\bibitem[{\citenamefont{Lohse et~al.}(2016)\citenamefont{Lohse, Schweizer,
  Zilberberg, Aidelsburger, and Bloch}}]{Lohse:2016aa}
\bibinfo{author}{\bibfnamefont{M.}~\bibnamefont{Lohse}},
  \bibinfo{author}{\bibfnamefont{C.}~\bibnamefont{Schweizer}},
  \bibinfo{author}{\bibfnamefont{O.}~\bibnamefont{Zilberberg}},
  \bibinfo{author}{\bibfnamefont{M.}~\bibnamefont{Aidelsburger}},
  \bibnamefont{and} \bibinfo{author}{\bibfnamefont{I.}~\bibnamefont{Bloch}},
  \bibinfo{journal}{Nat Phys} \textbf{\bibinfo{volume}{12}},
  \bibinfo{pages}{350} (\bibinfo{year}{2016}).

\bibitem[{\citenamefont{Marra and Nitta}(2020)}]{Marra:2020wh}
\bibinfo{author}{\bibfnamefont{P.}~\bibnamefont{Marra}} \bibnamefont{and}
  \bibinfo{author}{\bibfnamefont{M.}~\bibnamefont{Nitta}},
  \bibinfo{journal}{Physical Review Research} \textbf{\bibinfo{volume}{2}},
  \bibinfo{pages}{042035} (\bibinfo{year}{2020}).

\bibitem[{\citenamefont{Nakajima et~al.}(2021)\citenamefont{Nakajima, Takei,
  Sakuma, Kuno, Marra, and Takahashi}}]{Nakajima:2021xw}
\bibinfo{author}{\bibfnamefont{S.}~\bibnamefont{Nakajima}},
  \bibinfo{author}{\bibfnamefont{N.}~\bibnamefont{Takei}},
  \bibinfo{author}{\bibfnamefont{K.}~\bibnamefont{Sakuma}},
  \bibinfo{author}{\bibfnamefont{Y.}~\bibnamefont{Kuno}},
  \bibinfo{author}{\bibfnamefont{P.}~\bibnamefont{Marra}}, \bibnamefont{and}
  \bibinfo{author}{\bibfnamefont{Y.}~\bibnamefont{Takahashi}},
  \bibinfo{journal}{Nature Physics}  (\bibinfo{year}{2021}).

\bibitem[{\citenamefont{Rice and Mele}(1982)}]{Rice:1982qf}
\bibinfo{author}{\bibfnamefont{M.~J.} \bibnamefont{Rice}} \bibnamefont{and}
  \bibinfo{author}{\bibfnamefont{E.~J.} \bibnamefont{Mele}},
  \bibinfo{journal}{Phys. Rev. Lett.} \textbf{\bibinfo{volume}{49}},
  \bibinfo{pages}{1455} (\bibinfo{year}{1982}).

\bibitem[{\citenamefont{Xiao et~al.}(2010)\citenamefont{Xiao, Chang, and
  Niu}}]{Xiao:2010fk}
\bibinfo{author}{\bibfnamefont{D.}~\bibnamefont{Xiao}},
  \bibinfo{author}{\bibfnamefont{M.-C.} \bibnamefont{Chang}}, \bibnamefont{and}
  \bibinfo{author}{\bibfnamefont{Q.}~\bibnamefont{Niu}},
  \bibinfo{journal}{Reviews of Modern Physics} \textbf{\bibinfo{volume}{82}},
  \bibinfo{pages}{1959} (\bibinfo{year}{2010}).

\bibitem[{\citenamefont{Streda}(1982)}]{Streda:1982aa}
\bibinfo{author}{\bibfnamefont{P.}~\bibnamefont{Streda}},
  \bibinfo{journal}{Journal of Physics C: Solid State Physics}
  \textbf{\bibinfo{volume}{15}}, \bibinfo{pages}{L717} (\bibinfo{year}{1982}).

\bibitem[{\citenamefont{Hatsugai and Fukui}(2016)}]{Hatsugai:2016aa}
\bibinfo{author}{\bibfnamefont{Y.}~\bibnamefont{Hatsugai}} \bibnamefont{and}
  \bibinfo{author}{\bibfnamefont{T.}~\bibnamefont{Fukui}},
  \bibinfo{journal}{Physical Review B} \textbf{\bibinfo{volume}{94}},
  \bibinfo{pages}{041102} (\bibinfo{year}{2016}).

\bibitem[{\citenamefont{Thouless et~al.}(1982)\citenamefont{Thouless, Kohmoto,
  Nightingale, and den Nijs}}]{Thouless:1982uq}
\bibinfo{author}{\bibfnamefont{D.~J.} \bibnamefont{Thouless}},
  \bibinfo{author}{\bibfnamefont{M.}~\bibnamefont{Kohmoto}},
  \bibinfo{author}{\bibfnamefont{M.~P.} \bibnamefont{Nightingale}},
  \bibnamefont{and} \bibinfo{author}{\bibfnamefont{M.}~\bibnamefont{den Nijs}},
  \bibinfo{journal}{Physical Review Letters} \textbf{\bibinfo{volume}{49}},
  \bibinfo{pages}{405} (\bibinfo{year}{1982}).

\bibitem[{\citenamefont{Harper}(1955)}]{Harper:1955aa}
\bibinfo{author}{\bibfnamefont{P.~G.} \bibnamefont{Harper}},
  \bibinfo{journal}{Proceedings of the Physical Society. Section A}
  \textbf{\bibinfo{volume}{68}}, \bibinfo{pages}{874} (\bibinfo{year}{1955}).

\bibitem[{\citenamefont{Fukui et~al.}(2005)\citenamefont{Fukui, Hatsugai, and
  Suzuki}}]{FHS05}
\bibinfo{author}{\bibfnamefont{T.}~\bibnamefont{Fukui}},
  \bibinfo{author}{\bibfnamefont{Y.}~\bibnamefont{Hatsugai}}, \bibnamefont{and}
  \bibinfo{author}{\bibfnamefont{H.}~\bibnamefont{Suzuki}},
  \bibinfo{journal}{Journal of the Physical Society of Japan}
  \textbf{\bibinfo{volume}{74}}, \bibinfo{pages}{1674} (\bibinfo{year}{2005}).

\bibitem[{\citenamefont{Hatsugai et~al.}(2006)\citenamefont{Hatsugai, Fukui,
  and Aoki}}]{Hatsugai:2006aa}
\bibinfo{author}{\bibfnamefont{Y.}~\bibnamefont{Hatsugai}},
  \bibinfo{author}{\bibfnamefont{T.}~\bibnamefont{Fukui}}, \bibnamefont{and}
  \bibinfo{author}{\bibfnamefont{H.}~\bibnamefont{Aoki}},
  \bibinfo{journal}{Physical Review B} \textbf{\bibinfo{volume}{74}},
  \bibinfo{pages}{205414} (\bibinfo{year}{2006}).

\end{thebibliography}

\end{document}